\documentclass[preprint,12pt]{elsarticle}
\usepackage[english]{babel}
\usepackage{caption}
\usepackage{fontenc}
\usepackage{subfigure}
\usepackage{amsmath}
\usepackage{subeqnarray}
\usepackage[all]{xy}
\usepackage{ctable}
\usepackage{stfloats}
\usepackage{dcolumn}
\usepackage{array}
\usepackage{amssymb}
\usepackage[version=3]{mhchem}
\usepackage{bbding}
\usepackage{longtable}
\usepackage{ifsym}
\usepackage{natbib}
\usepackage{amssymb}
\usepackage{hyperref}
\biboptions{sort&compress}

\graphicspath{{./img/}}

\usepackage{graphicx,color,psfrag}

\usepackage[letterpaper, margin=1in]{geometry}

\catcode`\@=11
\def\gtsim{\mathrel{\vcenter{\m@th\offinterlineskip
\hbox{$\hfill>\hfill$}\kern.5ex\hbox{$\hfill\sim\hfill$}}}}
\catcode`\@=12

\catcode`\@=11
\def\ltsim{\mathrel{\vcenter{\m@th\offinterlineskip
\hbox{$\hfill<\hfill$}\kern.5ex\hbox{$\hfill\sim\hfill$}}}}
\catcode`\@=12
\catcode`\@=12

\begin{document}
\baselineskip 18pt
\title{Minimum Ignition Energy of Methanol-Air Mixtures}

\author[uc3m]{Eduardo Fern\'andez-Tarrazo}
\author[uc3m]{Mario S\'anchez-Sanz}
\author[ucsd]{Antonio L. S\'anchez}
\author[ucsd]{Forman A. Williams}
\journal{Combustion and Flame}

\address[uc3m]{Dept. Ingenier\'{\i}a T\'ermica y de Fluidos, Universidad Carlos III de Madrid, Legan\'es 28911, Spain}
\address[ucsd]{Dept. Mechanical and Aerospace Engineering, University of California San Diego, La Jolla CA 92093-0411, USA}





\begin{abstract}

A method for computing minimum ignition energies for gaseous fuel mixtures with detailed and reduced chemistry, by numerical integration of time-dependent conservation equations in a spherically symmetrical configuration, is presented and discussed, testing its general characteristics and accuracy. The method is applied to methanol-air mixtures described by a 38-step Arrhenius chemistry description and by an 8-step chemistry description based on steady-state approximations for reaction intermediaries. Comparisons of predictions with results of available experimental measurements produced reasonable agreements and supported both the robustness of the computational method and the usefulness of the 8-step reduction in achieving accurate predictions. 

\end{abstract}


\maketitle


Reduced chemistry has recently been derived for methanol combustion, identifying 38 elementary steps that are sufficient and employing them to define a satisfactory 8-step systematically reduced mechanism, obtained through the introduction of suitable steady-state approximations \cite{FSSW2016}. The mechanism was tested against burning-velocity, diffusion-flame, and autoignition experiments, but not against measurements of minimum ignition energies (MIE's). Indeed, it is not usual to test predictions of mechanisms against MIE measurements because programs for MIE calculations are not readily available. In the present work, a computational approach based on an appropriate formulation for such tests is developed, and reasonable agreement between the predictions and measurements is found. While the agreement may not be surprising in view of the earlier degrees of success of the mechanism, the formulation itself, and its evaluation, as reported here, may be of more general interest, since it can readily be applied to other fuels, as well. 

The equations for conservation of mass, energy, and chemical species in spherical coordinates are

\begin{align}
  \frac{\partial \rho}{\partial t} + \frac{1}{r^2} \frac{\partial}{\partial r} \big[ r^2 \rho v \big] & = 0 \label{eq:mass}\\
  \frac{\partial \rho h}{\partial t} + \frac{1}{r^2} \frac{\partial}{\partial r} \big[ r^2 \rho v h \big] & = -\frac{1}{r^2} \frac{\partial }{\partial r} \big[ r^2 q \big] \label{eq:energy} \\
  {\color{black} \frac{\partial \rho Y_{\rm{ i}}}{\partial t} + \frac{1}{r^2} \frac{\partial}{\partial r} \big[ r^2 \rho v Y_{\rm{i}} \big] }& = -\frac{1}{r^2} \frac{\partial }{\partial r} \big[ r^2 J_{\rm{i}} \big] + \omega_{\rm{i}} 
  \label{eq:Y}
\end{align}
supplemented with the isobaric equation of state,
\begin{equation}
  \frac{\rho T}{W} = \frac{\rho_0 T_0}{W_0},
\end{equation}
where the heat flux is given by the generalized Fourier law, $q = -k \partial T /\partial r + \Sigma J_{\rm{i}} h_{\rm{i}}$, and the molecular diffusive fluxes are described by Fick's law $J_{\rm{i}} = -\rho D_{\rm{i}} \partial Y_{\rm{i}}/\partial r$, with conductivity $k$ and diffusivities $D_{\rm{i}}$ given in \cite{SmGi}, extended in \cite{FSSW2016} to include methanol, producing excellent results for premixed flames. Here $W$ is the molecular mass of the mixture, obeying $1/W = \Sigma Y_{\rm{i}} /W_{\rm{i}}$, and $\omega_{\rm{i}}$ is the mass rate of production of species i.  The boundary conditions are
\begin{equation}
  \begin{array}{lcll}
   r=0          & : & \dfrac{\partial T}{\partial r} = 0;  & \dfrac{\partial Y_{\rm{i}}}{\partial r} = 0, \; \rm{i}=1, \dots N \\
   & & & \\
   r \to \infty & : & T=T_0;                              &    Y_{\rm{i}} = Y_{\rm{i}_0},  \; \rm{i}=1, \dots N.              \\ 
  \end{array}
\end{equation}


These equations apply to a uniform methanol-air mixture, initially at pressure $p_0$ and temperature $T_0$, of composition $Y_{\rm{i}_0}$. At $t=0$ a point source of heat with power {\color{black}$\dot Q$} is switched on during a time $t_d$, thereby providing an energy {\color{black}$E = \int_0^{t_{\rm{d}}} \dot Q {\rm d} t$} to the gas. From an energy balance, it is possible to relate the size of the resulting hot kernel $r_{\mathrm{h}}$ to the energy $E=4 \pi \rho_0 c_{p_{\rm{h}}} T_0 r_{\rm{h}}^3 /3$ as has been demonstrated in \cite{Kurdyumov2004}. The most favorable conditions for combustion initiation, leading to the minimum values of the ignition energy, are encountered when the deposition time $t_d$ is long compared with an acoustic time (defined as the ratio of $r_{\mathrm{h}}$ to the ambient sound speed) but short compared with a conduction time $t_c$ (defined as the ratio of $r_{\mathrm{h}}^2$ to the thermal diffusivity evaluated with $p_0$ and $T_0$). The potential singularity associated with a possible infinitely large temperature at the origin $r=0$ is avoided by using the formulation developed by Kurdyumov et al. \cite{Kurdyumov2004}, in which the integration is initiated at $t=t_0 \ll t_c$ using 
\begin{equation}
  \left.
    \begin{array}{lll}
    r < r_{\mathrm{h}}: & T=T_{\rm{h}}(r); &  Y_{\mathrm{i}} = Y_{\mathrm{i, h}}, \; \rm{i}=1, \dots N \\
    r > r_{\mathrm{h}}: & T=T_0;    & Y_{\mathrm{i}} = Y_{\mathrm{i, 0}}, \; \rm{i}=1, \dots N,
  \end{array}
  \right.
\end{equation}
where $T_{\rm{h}}$ and $Y_{\mathrm{i, h} }$ are temperature and mass-fraction distributions within the hot kernel, extended here to apply to methanol. As shown in \cite{Kurdyumov2004}, the initial temperature can be evaluated at $t=t_0$ using the self-similar solution 
{$\left[T_{\rm{h}}(r)-T_0\right]/T_0=(t_c/t_0)^{1/(\sigma+1)} \Theta(r/r_{\mathrm{h}})$, where $\sigma=0.7$} is the exponent for the presumed power-law temperature dependence of the thermal conductivity and $\Theta(r/r_{\mathrm{h}})$ is a self-similar function. The compositions inside the kernel $Y_{\mathrm{i,h}}$ can be obtained from major-species equilibrium as functions of the equivalence ratio $\phi$. For lean mixtures of methanol-air ($\phi < 1$),
\begin{equation}
  \frac{2}{3} \phi {\rm CH_3OH} + {\rm O_2} + 3.76 {\rm N_2} \rightarrow \frac{2}{3} \phi {\rm CO_2} + \frac{4}{3} \phi {\rm H_2O} + (1-\phi) {\rm O_2} + 3.76 {\rm N_2}.
\end{equation}
For rich mixtures ($\phi > 1$), on the other hand,
\begin{equation}
  \begin{array}{rl}
  \frac{2}{3} \phi {\rm CH_3OH} + {\rm O_2} + 3.76 {\rm N_2} 
  \rightarrow &
  3.76 {\rm N_2} + \frac{4}{3} \alpha \phi {\rm H_2} + \frac{4}{3} (1-\alpha) \phi {\rm H_2O} \\ 
              &
  + \left[ 2 - \frac{4}{3} (1-\alpha) \phi \right] {\rm CO_2} 
  + \left[ 2 (\phi-1) - \frac{4}{3} \alpha \phi \right] {\rm CO},
  \end{array}
\end{equation}
in which case the mixture composition depends on both $\phi$ and $\alpha$. While $\phi$ is fixed a priori, $\alpha$ is a parameter that gives the ratio of ${\rm H}_2$ to ${\rm H_2O}$ in the products. Although its evaluation in advance would require calculations of thermodynamic equilibrium, computations performed for the extreme values $\alpha=0$ and $\alpha=1$ showed that the actual value of $\alpha$ plays a negligible role in the determination of the minimum ignition energy. In fact, the actual initial temperatures in the hot kernel, of the order of $10^5$ K, make the reactions extremely fast, so that chemical equilibrium is reached essentially immediately after $t_0$; subsequently, in very short times, the temperature drops below 5000~K, and the composition of the mixture inside the hot kernel evolves rapidly to chemical equilibrium. Shortly afterward, usual flame temperatures are reached, and the chemical kinetics then begin to play a role in the evolution of the hot-gas kernel.  

The problem as formulated here is an evolution problem in which, for given initial conditions ($p_0$, $T_0$, and composition $Y_{\rm{i}_0}$), a given amount of energy $E$ released at the origin either succeeds in generating a self-propagating flame or results in the eventual termination of heat release. To determine which of these two histories occurs, a location $r_f$ is defined and tracked as the radial distance at which the heat-release rate is maximum. It was found that this radius either eventually grows with time or reaches a maximum and then decreases to zero. A slight decrease in the value of the energy parameter $E$ switches the evolution {from} the former behavior to the latter. The value at which this switch occurs is defined computationally as the MIE.

  %

\begin{center}
\begin{figure}[htp!]
\begin{center}
\includegraphics[width=.5\textwidth]{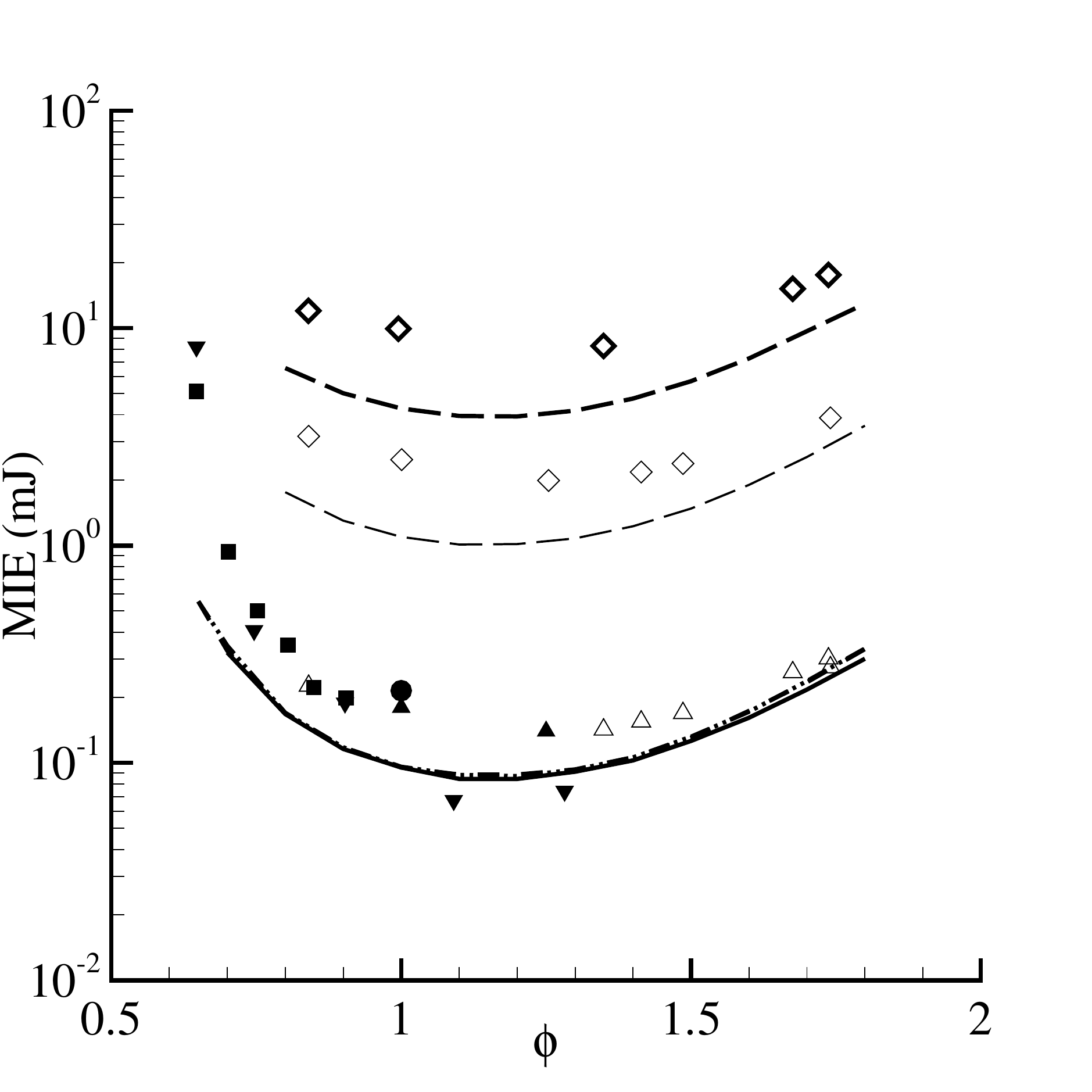} 
\caption{MIE as a function of equivalence ratio. Results for $p=1 \; \mathrm{bar}$, $T=300 \; \mathrm{K}$: {\large $\bullet$}  Calcote et al. \cite{Calcote52};  $\blacksquare$ Danis et al. \cite{Danis88};  $\blacktriangle$ Metzler \cite{Metzler53};   
$\triangle$  Metzler \cite{Metzler53} extrapolated in this work; { $\blacktriangledown$ Knepper \cite{Knepper2004}}; (Solid line) 38-steps; (Dashed line) 8-steps;
(Dash-dot-dotted line) 8-steps with point heat source $\dot q=E/t_{\rm{d}}$, $t_d=2 \times 10^{-5} ~ \mathrm{s}$. Results for $p=200$ mm Hg: {\large $\diamond$}  Metzler \cite{Metzler53};  (Thin long-dashed line) 8-steps; Results for $p=100$ mm Hg: {\boldmath{\large $\diamond$}}  Metzler \cite{Metzler53};  (Thick long-dashed line) 8-steps}
\label{fig:MIE_vs_phi}
\end{center}
\end{figure}
\end{center}

{ The differential equations were integrated by a second-order, finite-volume method, implicit in time, including tests to ensure space and time grid-independence. The chemical-kinetic rate parameters and thermodynamic properties were based on the San Diego mechanism, as previously described [1], with heat capacities set constant above 5000 K to prevent divergences that may develop from the NASA polynomial thermodynamic fits outside their range of validity.}
The calculated MIE values are plotted in Fig.~\ref{fig:MIE_vs_phi} as a function of equivalence ratio $\phi$ at ambient conditions for three different pressures. 
Calculations were carried out for two different energy-deposition histories, one starting the numerical integration at $t=0$ at ambient conditions and using a point heat source with constant power $E/t_{\rm{d}}$ for $t < t_{\rm{d}} = 2 \times 10^{-5} \, \mathrm{s}$, and the other using a self-similar initial condition [3] at $t_0=10^{-7} \, \mathrm{s}$. The results were indistinguishable, thereby supporting the robustness of the computational approach. It was found that predictions obtained with the multipurpose 8-step mechanism and with the 38-step mechanism \cite{FSSW2016} produced curves that lie on top of each other in the figure, thereby supporting the utility of the mechanism reduction.

Comparisons with experimental results obtained by Calcote et al. \cite{Calcote52} and Danis et al. \cite{Danis88} for lean to stoichiometric mixtures are seen in the figure to agree within a factor of 2, which is favorable, in view of the experimental difficulties discussed in those papers. Metzler \cite{Metzler53}, obtained results for $p=100$ and $p=200$ mm Hg and developed a method for extrapolating his results to ambient pressure, at stoichiometry, $\phi=1$, and also at $\phi=1.25$, where the minimum of the minimum ignition energy was found. It is noteworthy that this minimum occurs at the same equivalence ratio both experimentally and computationally. 
{In the present work, Metzler's extrapolation was found to be sufficiently accurate when compared with numerical simulations carried out for the same conditions, within this range of pressure.}
In Fig.~\ref{fig:MIE_vs_phi}, his extrapolations to atmospheric pressure are shown along with extrapolations from his results that were carried out in this work using the same pressure-dependence law 
{($MIE \sim p^{-2}$)}. These results, shown in the figure, exhibit good agreement with the results of the calculations, as well as with experimental results from other authors. The figure also shows the recent experimental results of Knepper \cite{Knepper2004}, exhibiting reasonable agreement, for lean mixtures, with evident disagreements between different experimental measurements for rich mixtures, illustrating difficulties associated with determination of fuel-rich minimum ignition energy for liquid fuels. 
{In all of these experiments, the period of energy deposition was large enough that shock-wave formation \cite{KSL} can be neglected. }

{In the surprisingly large temperature range between 300~K and 1000~K, the computational results are correlated extremely well by the formula $MIE = 0.24 \times 2.59^{-(T/300K)}$, where MIE is given in mJ, inspired by the findings reported in \cite{Calcote55}.} 
Above 1000~K autoignition effects become apparent in the computations; for levels of energy below the usual minimum ignition energy, the initial flame kernel was found to extinguish, but the region of hot gases remaining produced self-ignition within times of the order of $t_{\rm{m}}$, the ratio of a thermal diffusivity to the square of the laminar burning velocity (a measure of the transit time through a laminar flame) resulting in a self-propagating flame that persisted. These MIE results for methanol serve to illustrate how successful computations can be designed and completed for prevaporized liquid fuels of practical interest. 





\section*{Acknowledgements}

This work was supported by the Spanish MCINN through projects \# CSD2010-00011, ENE2012-33213 and ENE2015-65852-C2-1-R.

\section*{References}

\bibliographystyle{elsart-num}

\end{document}